\documentclass[doublecol]{epl2}
\usepackage{amsmath}
\usepackage{graphicx}
\usepackage{notoccite}

\shorttitle{Title}
\shortauthor{E. Ding \etal}
\institute{
  \inst{1} Department of Mathematics and Physics, Azusa Pacific University, Azusa, CA 91702-7000, USA\\
  \inst{2} Department of Mechanical Engineering, University of Hong Kong, Pokfulam Road, Hong Kong\\
  \inst{3} School of Engineering, Fraser Noble Building, King's College, University of Aberdeen, Aberdeen AB24 3UE, United Kingdom\\
  \inst{4} Department of Physical Electronics, School of Electrical Engineering, Faculty of Engineering, Tel Aviv University, Tel Aviv 69978, Israel\\
  \inst{5} Laboratory of Nonlinear Optical Informatics, ITMO University, St. Petersburg 197101, Russia
}
\abstract{We introduce a one-dimensional model based on the nonlinear Schr\"{o}dinger/Gross-Pitaevskii equation where the local nonlinearity is
subject to spatially periodic modulation in terms of the Jacobi $\mathrm{dn}$ function, with
three free parameters including the period, amplitude, and internal form-factor.
An exact periodic solution is found for each set of parameters and, which is more important for physical
realizations, we solve the inverse problem and predict
the period and amplitude of the modulation that yields a particular exact spatially periodic state.
Numerical stability analysis demonstrates that
the periodic states become modulationally unstable for large periods, and regain stability in the limit of an infinite period, which
corresponds to a bright soliton pinned to a localized nonlinearity-modulation pattern. Exact dark-bright
soliton complex in a coupled system with a localized modulation structure is also briefly considered . The
system can be realized in planar optical waveguides and cigar-shaped atomic Bose-Einstein condensates.}
\begin{document}

\title{Exact states in waveguides with periodically modulated nonlinearity}
\author{E. Ding\inst{1} \and H. N. Chan\inst{2} \and K. W. Chow\inst{2} \and %
K. Nakkeeran\inst{3} \and B. A. Malomed\inst{4,5}}
\maketitle

\section{Introduction}


It is commonly known that optical spatial solitons arise in planar and bulk
waveguides through the balance of the Kerr nonlinearity and transverse
diffraction~\cite{agrawal}. Modern fabrication technologies make it possible
to create waveguides featuring spatially inhomogeneous nonlinearities that
support novel classes of propagation patterns~\cite{boris1}. In particular,
spatially inhomogeneous waveguides with a defocusing nonlinearity, whose
local strength grows toward the periphery, can support diverse species of
fundamental and higher-order solitons, including vortices, necklace rings,
vortex gyroscopes, \textit{hopfions}, and complex hybrid modes~\cite%
{boris2,Lei_Wu,wu, zhong,Radik,Yasha,hybrids}, as well as \textit{localized
dark solitons} \cite{Zeng}. Similar nonlinearity landscapes, featuring
different growth rates of the local nonlinearity in opposite transverse
directions, support strongly asymmetric bright solitons~\cite{boris4}.
Asymmetric solitons also appear spontaneously if the nonlinearity profile
features a dual-well structure \cite{dual,Nir}. Furthermore, a combination
of the fast growing local strength of the defocusing nonlinearity with the
usual $\mathcal{PT}$-symmetric gain-loss profile makes it possible to
produce solitons that exhibit \textit{unbreakable} $\mathcal{PT}$ symmetry
\cite{unbreakable,raju,2D}, which is essential for constructing robust
solitons in such systems \cite{Demetri,PTreview1,PTreview2}. It is also
relevant to mention that a combination of $\mathcal{PT}$-symmetric with
competing nonlinearities supports spatiotemporal solitons \cite{ref1}.

Considerable interest has also been drawn to models with uniform
nonlinearity, either self-defocusing or focusing, and specially designed
periodic potentials that support exact periodic wave solutions~\cite%
{Carr1,Carr2}. Although both the particular potentials and the corresponding
exact periodic solutions are not generic, and the analysis of their
stability can only be performed numerically, these models provide direct
insight into the possibility to support periodic wave patterns by utilizing
the interplay of periodic potentials and the ubiquitous cubic nonlinearity.
Furthermore, nontrivial exact solutions serve as benchmarks which suggest
the shape of generic solutions. The inverse problem, aimed at engineering
waveguiding potentials adjusted to maintaining periodic waves with
prescribed properties, is a physically relevant issue too \cite{Spain}.

In this work, we introduce a model with a class of periodic modulations that
represent spatially periodic \textit{pseudopotentials} \cite{pseudo} induced
by the local nonlinearity. This model admits exact solutions in the form of
periodic wave patterns which, in the limiting case of an infinite modulation
period, become bright solitons. Stability of these patterns is studied
numerically. The same model can also be used to solve the inverse problem of
engineering a nonlinearity-modulation profile needed to support a wave
pattern with prescribed period and amplitude.

\section{The Mathematical Model}

The light propagation in a planar waveguide with spatially modulated
nonlinearity is described by the model that is based on the scaled
nonlinear Schr\"{o}dinger equation for the electromagnetic wave amplitude $%
\Psi (x,z)$,
\begin{equation}
i\Psi _{z}+\Psi _{xx}+g(x)|\Psi |^{2}\Psi =0,  \label{eq:nls}
\end{equation}%
where $x$ and $z$ are the transverse and longitudinal coordinates,
respectively. The periodically-modulated nonlinearity profile is defined by $%
g(x)$, chosen as
\begin{equation}
g(x)=\frac{\alpha }{\mathrm{dn}^{2}(x)}+\beta +\gamma ~\mathrm{dn}^{2}(x),
\label{eq:g}
\end{equation}%
where $\alpha $, $\beta $, and $\gamma $ are real constants, and $\mathrm{dn}%
(x)$ is the standard Jacobi elliptic function with modulus $\sqrt{m}$ and
period $2K$ ($K$ being the complete elliptic integral of the first kind).
It is relevant to mention that, in the general case, the periodic
inhomogeneity affects not only the local nonlinearity, but also the local
refractive index, which would generate an additional term $U(x)\Psi $ in
Eq.~(\ref{eq:nls}), with an effective spatially-periodic
potential, $U(x)$. Nevertheless, specific experimental methods,
such as resonant doping, make it possible to create waveguides in which the
nonlinearity is affected by the periodic modulation, while the refractive
index remains nearly constant~\cite{boris1}.

The same model, with propagation distance $z$ replaced by time $t$,
represents the scaled Gross-Pitaevskii equation for the mean-field wave
function of an atomic Bose-Einstein condensate (BEC), for which the periodic
nonlinearity modulation can be induced by means of the Feshbach resonance in
a spatially non-uniform magnetic or optical field. In particular, the
necessary periodic profile of the magnetic field can be accurately
implemented by means of the known technique based on the use of
appropriately designed magnetic lattices~\cite{magnetic}. Furthermore, the
spatially periodic distribution of the local nonlinearity coefficient in BEC
has been experimentally realized by means of an optical lattice~\cite%
{Feshbach}. A particular anharmonic profile corresponding to Eq.~(\ref{eq:g}%
) can be effectively approximated by a superposition of several harmonics of
its Fourier decomposition, represented by the respective optical lattices.
In a real experiment, the setting also includes an overall parabolic
trapping potential. However, in many situations the characteristic scale of
this potential is much larger than the period of the spatial modulation,
which makes it possible to neglect the trapping potential while analyzing
the effects of periodic lattices~\cite{Heidelberg}.

We look for a stationary solution to Eq.~(\ref{eq:nls}) in the form of a
\textquotedblleft $\mathrm{dn}$-wave":
\begin{equation}
\Psi (x,z)\equiv \psi (x)e^{-i\Omega z}=\frac{A_{0}\;\mathrm{dn}(x)}{1+b\;%
\mathrm{dn}^{2}(x)}e^{-i\Omega z},  \label{eq:ansatz}
\end{equation}%
where $A_{0}$ and $b$ are real constants, and $-\Omega $ is the propagation
constant (or the chemical potential, in the BEC model). Note that this
ansatz is nonsingular under the conditions $b>-1$ or $b<-1/(1-m)$.
Substituting it into Eq.~(\ref{eq:nls}) gives rise to the following system
of equations for the three ansatz parameters $A_{0}^{2}$, $b$, and $\Omega $%
:
\begin{equation}
\left\{ \begin{aligned} 2+A_{0}^{2} \; \alpha -6b(m-1)-m+\Omega =0 \; , \\
A_{0}^{2} \; \beta +2b^{2}(m-1)+2b(\Omega +3m-6)-2=0 \; , \\ 6b+A_{0}^{2} \;
\gamma +b^{2}(2-m+\Omega )=0 \; . \end{aligned}\right.  \label{eq:para}
\end{equation}%
One can calculate any three parameters from this system for given values of
the others. In particular, this allows one to address the above-mentioned
inverse problem, aimed at determining the nonlinearity modulation profile
(see Eq.~(\ref{eq:g})) needed for maintaining a particular wave pattern.

\section{Stability Analysis}

\begin{figure}[t]
\begin{center}
\includegraphics[width = 75mm,keepaspectratio]{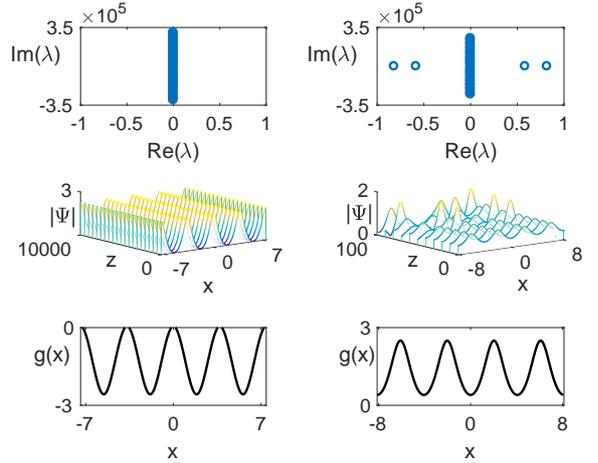}
\end{center}
\caption{Left: The eigenvalue spectrum (top), simulated propagation
(middle), and modulation profile (bottom) corresponding to a stable $\mathrm{%
dn}$-wave with $A_{0}^{2}=1$, $b=-0.6$, $m=0.5$, $\Omega =1.7$, $\protect%
\alpha =-1.4$, $\protect\beta =-1$, and $\protect\gamma =2.448$. Right: The
results for an unstable wave with $A_{0}^{2}=25.1989$, $b=-12.968$, $m=0.67$%
, $\Omega =-0.8523$, $\protect\alpha =1$, $\protect\beta =-0.5$, and $%
\protect\gamma =-0.1$.}
\label{fig:stab}
\end{figure}
The stability of the $\mathrm{dn}$-wave denoted by Eq.~(\ref{eq:ansatz}) is
investigated by means of the standard linearization procedure~\cite%
{linstab1,linstab2}. Substituting $\Psi (x,z)=\left[ \psi (x)+u(x,z)\right]
e^{-i\Omega z}$ into Eq.~(\ref{eq:nls}), with the small complex perturbation
defined as $u\left( x,z\right) \equiv R(x,z)+iI(x,z)$, we arrive at the
linearized system,
\begin{equation}
\left\{ \begin{aligned} \partial _{z}R &=&\left( -\Omega -g(x)\psi
^{2}(x)-\partial _{x}^{2}\right) I \; ,\\ \partial _{z}I &=&\left( \Omega
+3g(x)\psi ^{2}(x)+\partial _{x}^{2}\right) R \; . \end{aligned}\right.
\label{eq:linear}
\end{equation}%
The stability of the $\mathrm{dn}$-wave is determined by substituting $%
\left\{ R(x,z),I(x,z)\right\} =\left\{ P(x),Q(x)\right\} \exp (\lambda z)$
into the above equations. The resulting problem for stability eigenvalue $%
\lambda $ is solved numerically using the finite-difference method. In
particular, \textit{modulational instability} of periodic states~\cite%
{agrawal} is accounted for by eigenvalues with $\mathrm{Re}(\lambda )>0$.
Generic examples of stable and unstable $\mathrm{dn}$-waves are presented in
Fig.~\ref{fig:stab}. The stability, as predicted by the calculation of the
eigenvalue spectra, is corroborated by direct simulations of Eq.~(\ref%
{eq:nls}), using the Fourier transform in $x$ and a fourth-order Runge-Kutta
algorithm in $z$.

\begin{figure}[t]
\begin{center}
\includegraphics[width = 80mm,keepaspectratio]{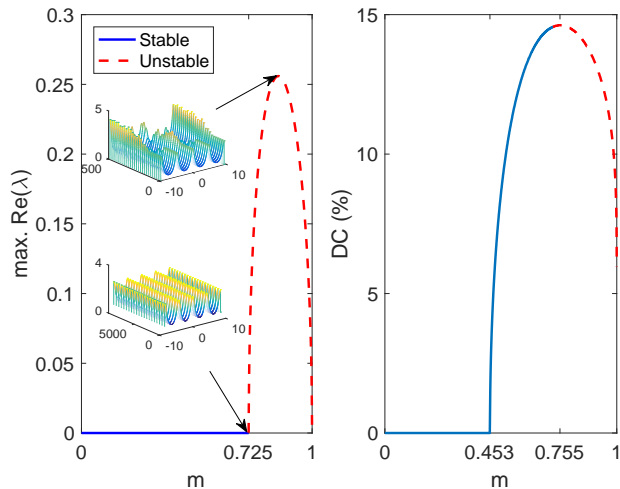}
\end{center}
\caption{Left: The variation of the largest real part in the eigenvalue
spectrum for the stationary \textrm{dn}-wave solution, as a function of the
squared elliptic modulus, $m$, while three other parameters are fixed as $%
A_{0}^{2}=1$, $b=-0.6$, and $\protect\beta =-1$. The top and bottom
three-dimensional subplots show the simulated evolution at $m=0.8518$ and $%
m=0.7240$, respectively. Right: The duty cycle ($\mathrm{DC}$) of $g(x)$
(defined in the main text) as a function of $m$. }
\label{fig:branch}
\end{figure}

The dependence of the stability of the $\mathrm{dn}$-waves on the system's
parameters can be explored with the help of numerical-continuation
techniques~\cite{Alan,cont,Doedel}. In particular, it is important to know
how the stability is affected by varying the nonlinearity-modulation period $%
2K$, which is determined by the squared modulus, $m$. We fix $A_{0}$, $b$,
and $\beta $ in Eq.~(\ref{eq:para}) as $A_{0}^{2}=1$, $b=-0.6$, and $\beta
=-1$, and determine the other parameters, \textit{viz}., $\alpha $, $\gamma $%
, and $\Omega $, for each value of $m$. The results are summarized in Fig.~%
\ref{fig:branch}. In this case, the $\mathrm{dn}$-wave is found to be stable
in the region of $0\leq m\leq 0.725$, where none of the eigenvalues in the
spectrum has a positive real part (see the left panel). For $m>0.725$, the
long-period $\mathrm{dn}$-waves are destabilized by at least one eigenvalue
with $\mathrm{Re}(\lambda )>0$. The strongest instability is found at around
$m=0.852$, where the $\mathrm{dn}$-wave is quickly destroyed by the
instability. The evolution of the wave profiles at the onset of the
instability, as well as the strongest-instability point, are also shown in
the left panel. The \textit{duty cycle} ($\mathrm{DC}$) of the modulation
profile, i.e., the share of the region carrying a self-focusing nonlinearity
per one period of $g(x)$, is shown in the right panel of Fig.~\ref%
{fig:branch}. The nonlinearity is entirely self-defocusing at $m<0.453$
where $\mathrm{DC}\equiv 0$. Once $m$ exceeds this threshold, the $\mathrm{DC%
}$ first increases to a maximum of $14.62\%$ at $m=0.755$, and then
approaches zero in the long-wave limit of $m\rightarrow 1$ where the
modulation period $2K$ becomes infinite. We have found that the mean value
of $g(x)$ is always negative in the entire range of $m$ values, i.e., the
nonlinearity is self-defocusing on average. It is worthy to note that the
maximum of the $\mathrm{DC}$ roughly coincides with the onset of instability.

In the long-wave limit of $m\rightarrow 1$, the modulation profile (Eq.~(\ref%
{eq:g})) assumes the localized shape:
\begin{equation}
g(x)=\alpha \cosh ^{2}x+\beta +\gamma \;\mathrm{sech}^{2}x\;.
\label{eq:sech}
\end{equation}%
The left panel of Fig.~\ref{fig:branch} suggests that the instability of the
\textrm{dn}-wave vanishes in this limit, with the corresponding stable exact
solution (see Eq.~(\ref{eq:ansatz})) being a bright soliton:
\begin{equation}
\Psi (x,z)=\frac{A_{0}\;\mathrm{sech}(x)}{1+b~\mathrm{sech}^{2}(x)}%
e^{-i\Omega z}\;.  \label{eq:soliton}
\end{equation}%
The stability of the soliton solution is analyzed here only for the case
where $\alpha =0$, to ensure that the localized modulation profile (Eq.~(\ref%
{eq:sech})) is not singular at $|x|\rightarrow \infty $ (nevertheless, the
self-defocusing singularity with $\alpha <0$ may readily support robust
self-trapped modes~\cite%
{boris2,Lei_Wu,wu,zhong,Radik,Yasha,hybrids,boris4,dual,Nir}). In this
situation, i.e., with $\alpha =0$ and $m\rightarrow 1$, system~(\ref{eq:para}%
) yields
\begin{equation}
\Omega =-1\;,\;A_{0}^{2}=\frac{2(1+4b)}{\beta }\;,\;\gamma =-\frac{3b\beta }{%
1+4b}\;.  \label{eq:reduced}
\end{equation}

The stability condition for the soliton pinned to the spatially modulated
nonlinearity profile given by Eq. (\ref{eq:sech}) with $\alpha =0$, is $%
\gamma >0$, as in that case the soliton is pulled to the local maximum of
self-attraction. Equations (\ref{eq:soliton}) and~(\ref{eq:reduced}) admit $%
\gamma >0$ in two cases:%
\begin{equation}
\beta >0,~0<-b<1/4 \; ;  \label{beta>0}
\end{equation}%
\begin{equation}
\beta <0,~1/4<-b<1 \; .  \label{beta<0}
\end{equation}%
In the former case, the nonlinearity is globally self-focusing, while in the
latter one a finite self-focusing region (``defect") is embedded in a
defocusing background.

\begin{figure}[t]
\begin{center}
\includegraphics[width = 80mm,keepaspectratio]{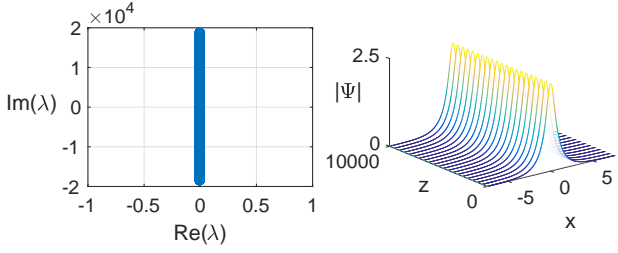}
\end{center}
\caption{Left: The eigenvalue spectrum for the soliton given in Eq.~(\protect
\ref{eq:soliton}) with $\protect\beta =-2.8$, $b=-0.6$, $\protect\gamma =3.6$%
, and $A_{0}^{2}=1$. Right: Stable evolution of the soliton.}
\label{fig:sech}
\end{figure}

\begin{figure}[t]
\begin{center}
\includegraphics[width = 75mm,keepaspectratio]{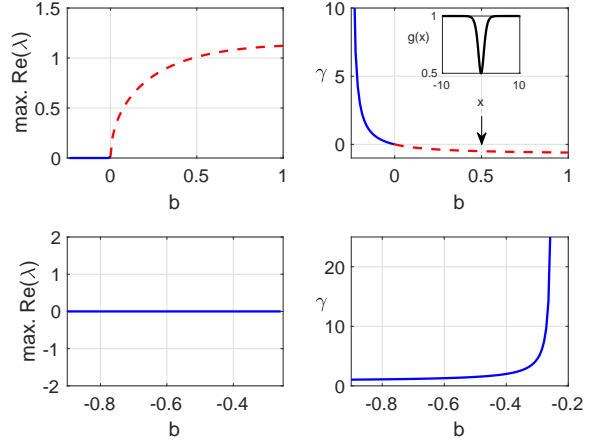}
\end{center}
\caption{Top: The largest real part in the eigenvalue spectrum of the
soliton (right) and parameter $\protect\gamma $ (left) as functions of $b$,
while $\protect\beta =1$ is fixed. Solid and dashed curves represent stable
and unstable soliton branches, respectively. The subplot within the top
right panel shows the modulation profile $g(x)$ at $b=0.5$. Bottom: The
corresponding results for $\protect\beta =-1$.}
\label{fig:soliton_stab}
\end{figure}

An example of the latter situation is shown in Fig.~\ref{fig:sech}, where
the stability of the pinned bright soliton is confirmed by both the
eigenvalue spectrum and direct simulations. In this case, $\beta =-2.8$ and $%
\gamma =3.6$,$\ g(x)$ being positive at $|x|<0.74$. In fact, stable solitons
can be produced in a wide range of parameter values, as shown in Fig.~\ref%
{fig:soliton_stab}. Unstable solitons are only found in the case of $\beta
>0>\gamma $ (represented by red dashed curves in the top panels). In this
case, the self-focusing is stronger farther from the center, hence the
soliton is repelled by the effective nonlinear potential. An example of such
a nonlinearity profile is shown in the subplot in the top right panel of
Fig.~\ref{fig:soliton_stab}.

\begin{figure}[t]
\begin{center}
\includegraphics[width = 75mm,keepaspectratio]{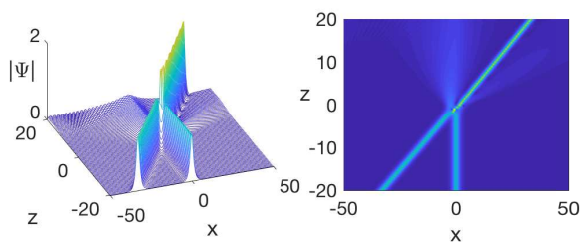}
\end{center}
\caption{The collision of a free bright soliton, given by Eq. \protect\ref%
{free-bright} with parameters $\protect\beta = 1$, $c = 1.65$, $\protect\eta %
= 1$, and a pinned one with $\protect\beta = 1$, $b = -0.1$, $\protect\gamma %
= 0.5$, and $A_0^2 = 1.2$.}
\label{fig:collision}
\end{figure}

Collisions between solitons play an important role in the study of their
dynamics. In the case corresponding to condition~(\ref{beta>0}), one may
consider the collision of a free bright soliton, with inverse width $\eta $
and velocity (slope) $c$,%
\begin{equation}
\Psi _{\mathrm{free}}\left( x,z\right) = \sqrt{\frac{2}{\beta}} \eta\;%
\mathrm{sech}\left( \eta \left( x-cz\right) \right) e^{\left( \frac{i}{2}%
cx+i\left( \eta ^{2}-\frac{c^{2}}{4}\right) z\right)} \; ,
\label{free-bright}
\end{equation}%
with a pinned soliton. An example of such a collision is displayed in Fig.~%
\ref{fig:collision}. The incident soliton captures the pinned one, merging
with it into a single soliton which continues to move with original
velocity. This outcome may find applications to the design of soliton=based
data-processing schemes.
In the case of $\gamma \ll \beta $, i.e., $-b\ll 1/4$ (see Eqs. (\ref%
{eq:reduced}) and (\ref{beta>0})), the collision can be considered by means
of the perturbation theory \cite{RMP}, which uses the exact result for the
collision-induced soliton's shift generated by the solution of the nonlinear
Schr\"{o}dinger equation. In this case, one may expects that the incident
soliton will pass through, while the pinned one will start oscillating
around the attractive nonlinear defect.

A completely novel situation arises in the case of Eq. (\ref{beta<0}), when
the model admits a freely moving dark soliton far from the defect. Its
collision with the pinned bright soliton will be governed by the repulsive
interaction, which may lead to various outcomes, such as rebound of the
incident dark soliton and destruction of the bright one through its
dislodgment from the pinned position. These possibilities call for
systematic numerical simulations of the collisions, which is a subject for a
separate work.

\section{The Manakov System}

Lastly, Eq.~(\ref{eq:nls}) can be generalized to a coupled system
\begin{equation}
\left\{ \begin{aligned} i\Psi _{z}+\Psi _{xx}+g(x)\left( |\Psi |^{2}+|\Phi
|^{2}\right) \Psi &=0 \;,\\ i\Phi _{z}+\Phi _{xx}+g(x)\left( |\Psi
|^{2}+|\Phi |^{2}\right) \Phi &=0 \end{aligned}\right.   \label{eq:coupled}
\end{equation}%
that describes the copropagation of light modes with orthogonal
polarizations in a bimodal waveguide, under the Manakov's condition that the
self-phase- and cross-phase-modulation coefficients are equal~\cite%
{Manakov,Menyuk}, as well as a binary Bose-Einstein condensate composed of
two hyperfine atomic states~\cite{binary-BEC} (in the latter case, the
relative nonlinearity is very close to the Manakov's point). In the
long-wave limit similar to Eq.~(\ref{eq:sech}) where $g(x)=\beta +\gamma ~%
\mathrm{sech}^{2}(rx)$, with $\gamma >0$ and parameter $r$
which determines the width of the attracting region, an exact solution of
the coupled equations can be found in the form of a stable \textit{symbiotic}
dark-bright soliton complex~\cite{symbio1,symbio2}:
\begin{equation}
\begin{aligned} \Psi (x,z) &=A_{0}\tanh (rx)e^{-i\Omega _{1}z} \;, \\ \Phi
(x,z) &=A_{0}\; \mathrm{sech}(rx)e^{-i\Omega _{2}z}\;, \end{aligned}
\label{eq:ms}
\end{equation}%
with
\begin{equation*}
A_{0}=\sqrt{\frac{2}{\gamma }}\;r\;,\;\Omega _{1}=-\frac{2r^{2}\beta }{%
\gamma }\;,\;\Omega _{2}=-r^{2}-\frac{2r^{2}\beta }{\gamma }\;.
\end{equation*}%
In this case, the dark component $\Psi $ cannot exist without the
interaction with the bright counterpart $\Phi $, and the background
supporting the dark component is modulationally\ stable when $\beta <0$. An
example of a stable dark-bright complex is displayed in Fig.~\ref%
{fig:darkbright}.
\begin{figure}[t]
\begin{center}
\includegraphics[width = 80mm,keepaspectratio]{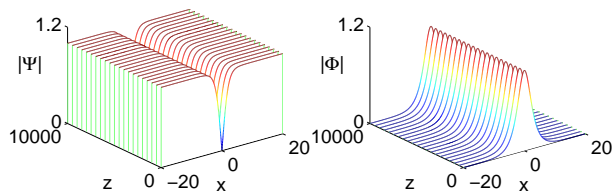}
\end{center}
\caption{The evolution of the stable dark-bright soliton complex produced by
Eq.~(\protect\ref{eq:ms}) for $\protect\beta =-1$, $\protect\gamma =0.5$, $%
r=0.5$, and $A_{0}=1$.}
\label{fig:darkbright}
\end{figure}

\section{Conclusion}

In this work, we have studied the one-dimensional model for the wave
transmission in a medium with a periodically-modulated local nonlinearity
that is based on the Jacobi elliptic \textrm{dn} function. The model, which
can be realized in optics and BEC~\cite{ref2}, admits both exact periodic
solutions and bright solitons (in the long-wave limit). Stable solutions of
these types provide a benchmark suggesting the shape of generic solutions
that can be found numerically in the same model. The model also allows for
the prediction of the modulation profile needed to support a particular
periodic wave form with prescribed period and amplitude. The numerical
analysis of the modulational stability has demonstrated that the periodic
patterns can be unstable for sufficiently large periods. However, stability
is retrieved in the limit of an infinite period which corresponds to bright
solitons. In addition, we have found an exact solution for dark-bright
soliton bound states in a similar two-component model that applies to
periodically inhomogeneous bimodal planar optical waveguides and binary BEC.
As an extension of the analysis, it will be relevant to study the periodic
solutions, soliton solutions, localized structures~\cite{ref3}, and in
particular bright-antidark soliton complex supported by this coupled model
in detail. Also, as mentioned above, it may be interesting to systematically
simulate collisions of a moving free bright or dark soliton with the pinned
one.

\acknowledgments
Partial financial support has been provided by the Research Grants Council
(Hong Kong) contract HKU 17200815.

\bibliography{ref}

\begin{thebibliography}{10}
\expandafter\ifx\csname url\endcsname\relax\def\url#1{\texttt{#1}}\fi

\bibitem{agrawal}
\Name{Kivshar Y.~S. \and Agrawal G.~P.} \Book{Optical solitons: From fibers to
  photonic crystals} (Academic Press) 2003.

\bibitem{boris1}
\Name{Kartashov Y.~V., Malomed B.~A. \and Torner L.} \REVIEW{Rev. Mod.
  Phys.}{83}{2011}{247}.

\bibitem{boris2}
\Name{Borovkova O.~V., Kartashov Y.~V., Torner L. \and Malomed B.~A.}
  \REVIEW{Phys. Rev. E}{84}{2011}{035602}.

\bibitem{Lei_Wu}
\Name{Tian Q., Wu L., Zhang Y. \and Zhang J.-F.} \REVIEW{Phys. Rev.
  E}{85}{2012}{056603}.

\bibitem{wu}
\Name{Wu Y., Xie Q., Zhong H., Wen L. \and Hai W.} \REVIEW{Phys. Rev.
  A}{87}{2013}{055801}.

\bibitem{zhong}
\Name{Zhong W.~P. \and Beli{\'c} M.} \REVIEW{Annals of Physics}{351}{2014}{787
  }.

\bibitem{Radik}
\Name{Driben R., Kartashov Y.~V., Malomed B.~A., Meier T. \and Torner L.}
  \REVIEW{Phys. Rev. Lett.}{112}{2014}{020404}.

\bibitem{Yasha}
\Name{Kartashov Y.~V., Malomed B.~A., Shnir Y. \and Torner L.} \REVIEW{Phys.
  Rev. Lett.}{113}{2014}{264101}.

\bibitem{hybrids}
\Name{Driben R., Kartashov Y.~V., Malomed B.~A., Meier T. \and Torner L.}
  \REVIEW{New Journal of Physics}{16}{2014}{063035}.

\bibitem{Zeng}
\Name{Zeng J. \and Malomed B.~A.} \REVIEW{Phys. Rev. E}{95}{2017}{052214}.

\bibitem{boris4}
\Name{Kartashov Y.~V., Lobanov V.~E., Malomed B.~A. \and Torner L.}
  \REVIEW{Opt. Lett.}{37}{2012}{5000}.

\bibitem{dual}
\Name{Xie Q., Wang L., Wang Y., Shen Z. \and Fu J.} \REVIEW{Phys. Rev.
  E}{90}{2014}{063204}.

\bibitem{Nir}
\Name{Dror N. \and Malomed B.~A.} \REVIEW{Journal of Optics}{18}{2016}{014003}.

\bibitem{unbreakable}
\Name{Kartashov Y.~V., Malomed B.~A. \and Torner L.} \REVIEW{Opt.
  Lett.}{39}{2014}{5641}.

\bibitem{raju}
\Name{Raju T.~S., Hegde T.~A. \and Kumar C.~N.} \REVIEW{J. Opt. Soc. Am.
  B}{33}{2016}{35}.

\bibitem{2D}
\Name{Guo D., Xiao J., Gu L., Jin H. \and Dong L.} \REVIEW{Physica D: Nonlinear
  Phenomena}{343}{2017}{1 }.

\bibitem{Demetri}
\Name{El-Ganainy R., Makris K.~G., Christodoulides D.~N. \and Musslimani Z.~H.}
  \REVIEW{Opt. Lett.}{32}{2007}{2632}.

\bibitem{PTreview1}
\Name{Suchkov S.~V., Sukhorukov A.~A., Huang J., Dmitriev S.~V., Lee C. \and
  Kivshar Y.~S.} \REVIEW{Laser \& Photonics Reviews}{10}{2016}{177}.

\bibitem{PTreview2}
\Name{Konotop V.~V., Yang J. \and Zezyulin D.~A.} \REVIEW{Rev. Mod.
  Phys.}{88}{2016}{035002}.

\bibitem{ref1}
\Name{Xu S.-L., Zhao Y., Petrovi{\'c} N.~Z. \and Beli{\'c} M.~R.} \REVIEW{EPL
  (Europhysics Letters)}{115}{2016}{14006}.

\bibitem{Carr1}
\Name{Carr L.~D., Clark C.~W. \and Reinhardt W.~P.} \REVIEW{Phys. Rev.
  A}{62}{2000}{063610}.

\bibitem{Carr2}
\Name{Carr L.~D., Clark C.~W. \and Reinhardt W.~P.} \REVIEW{Phys. Rev.
  A}{62}{2000}{063611}.

\bibitem{Spain}
\Name{Belmonte-Beitia J., P\'erez-Garc\'{\i}a V.~M., Vekslerchik V. \and
  Konotop V.~V.} \REVIEW{Phys. Rev. Lett.}{100}{2008}{164102}.

\bibitem{pseudo}
\Name{Calvayrac F., Reinhard P.-G., Suraud E. \and Ullrich C.} \REVIEW{Physics
  Reports}{337}{2000}{493 }.

\bibitem{magnetic}
\Name{Jose S., Surendran P., Wang Y., Herrera I., Krzemien L., Whitlock S.,
  McLean R., Sidorov A. \and Hannaford P.} \REVIEW{Phys. Rev.
  A}{89}{2014}{051602}.

\bibitem{Feshbach}
\Name{Yamazaki R., Taie S., Sugawa S. \and Takahashi Y.} \REVIEW{Phys. Rev.
  Lett.}{105}{2010}{050405}.

\bibitem{Heidelberg}
\Name{Morsch O. \and Oberthaler M.} \REVIEW{Rev. Mod. Phys.}{78}{2006}{179}.

\bibitem{linstab1}
\Name{Farnum E.~D. \and Kutz J.~N.} \REVIEW{J. Opt. Soc. Am.
  B}{25}{2008}{1002}.

\bibitem{linstab2}
\Name{Malomed B.~A., Ding E., Chow K.~W. \and Lai S.~K.} \REVIEW{Phys. Rev.
  E}{86}{2012}{036608}.

\bibitem{Alan}
\Name{Champneys A.} \REVIEW{Physica D: Nonlinear Phenomena}{112}{1998}{158 }
  proceedings of the Workshop on Time-Reversal Symmetry in Dynamical Systems.

\bibitem{cont}
\Name{Dhooge A., Govaerts W. \and Kuznetsov Y.~A.} \REVIEW{ACM Trans. Math.
  Softw.}{29}{2003}{141}.

\bibitem{Doedel}
\Name{Krauskopf B., Osinga H.~M., Doedel E.~J., Henderson M.~E., Guckenheimer
  J., Vladimirsky A., Dellnitz M. \and Junge O.} \REVIEW{International Journal
  of Bifurcation and Chaos}{15}{2005}{763}.

\bibitem{RMP}
\Name{Kivshar Y.~S. \and Malomed B.~A.} \REVIEW{Rev. Mod.
  Phys.}{61}{1989}{763}.

\bibitem{Manakov}
\Name{Manalov S.~V.} \REVIEW{Sov. Phys. JETP}{38}{1974}{248}.

\bibitem{Menyuk}
\Name{Menyuk C.~R.} \REVIEW{IEEE Journal of Quantum
  Electronics}{25}{1989}{2674}.

\bibitem{binary-BEC}
\Name{Ho T.-L.} \REVIEW{Phys. Rev. Lett.}{81}{1998}{742}.

\bibitem{symbio1}
\Name{P\'erez-Garc\'{\i}a V.~M. \and Beitia J.~B.} \REVIEW{Phys. Rev.
  A}{72}{2005}{033620}.

\bibitem{symbio2}
\Name{Adhikari S.~K.} \REVIEW{Physics Letters A}{346}{2005}{179 }.

\bibitem{ref2}
\Name{Das P., Noh C. \and Angelakis D.~G.} \REVIEW{EPL (Europhysics
  Letters)}{103}{2013}{34001}.

\bibitem{ref3}
\Name{Ding Y., Zhang B., Feng Q., Tang X., Liu Z., Chen Z. \and Lin C.}
  \REVIEW{EPL (Europhysics Letters)}{117}{2017}{14003}.

\end{thebibliography}
\bibliographystyle{eplbib}

%
%
%
%

\end{document}